%% file: gpv-rmcomment.tex
\documentclass[aps,prl,nofootinbib,nobibnotes,floatfix,twocolumn,showpacs,superscriptaddress]{revtex4-1}

\usepackage{pstricks}
\usepackage{graphicx}
\usepackage{bm}% bold math
\usepackage{amsmath,amssymb,amsfonts}
\usepackage{diagbox}

\usepackage[sort&compress]{natbib}

\let\vec\mathbf

\begin{document}

\title{Evidence for parity violation in gravitational fields}
\input{cauthors}

\begin{abstract}
\noindent
 Discrete symmetries in gravity have only been tested for
low energy, non-relativistic matter, confirming the perfectly symmetric
general relativity.
A hint for high energy $\cal{CP}$ violation in gravitational fields has recently been found 
in the HERA Compton polarimeter's two spectra, measured with electron and positron 
beams. Here we report results of the analysis of the same polarimeter's 314896 spectra, 
acquired during 2004--2007 and tagged by laser polarization
states allowing the separation of charge ($\cal C$) and space parity ($\cal P$) contributions.
The measured Compton edge energy asymmetry, induced by the laser helicity 
 flips, is as high as \hbox{$(4.9\pm0.5)\cdot10^{-5}$}
which corresponds to a helicity-dependent difference in the gravitational potentials
of \hbox{$(1.7\pm0.2)\cdot10^{-14}$}.
In the case of the observed anomalous coupling's energy independence, 
the spin asymmetric gravity will contribute to the galactic rotational curves. 
Further analysis and calculations can determine whether the observed magnitude of 
the gravitational parity violation is sufficient for 
detaching this famous phenomena from the dark matter theory.
\end{abstract}


\maketitle

\section{Introduction}

General Relativity (GR)~\cite{Einstein-GR}, the currently accepted 
theory of gravitation, rests on the principle of equivalence, which 
is a concept originating from the universality of free fall for massive 
bodies.
Up to now all macroscopic scale experiments have confirmed this 
basic gravitational principle down to the contemporary 
limit of $10^{-13}$ ~\cite{Will:2014kxa,Patrignani:2016xqp}.
In the microscopic realm of quantum and high-energy physics 
gravitation is completely negligible in comparison to the nuclear or electroweak 
forces between the particles.
Also the influence of gravity as an external field is largely ignorable for
quantum particle interactions.
Indeed, according to the equivalence principle, 
particles  with different
natures, intrinsic, spatial, or energetic properties
are affected (accelerated) equally by gravity.
 Such absolute democracy and symmetry makes any gravitational field undetectable 
for high energy particle interactions taking
place in a neighborhood small enough compared to the local curvature
radius of spacetime.

The situation is different if the symmetry is broken.
Then an instantaneous momentum exchange with the background gravitational 
field will depend on the particle's type and features affecting the quantum processes' 
kinematics and dynamics.
We are searching for such asymmetries in gravity, motivated by particle physics'
well known observations that the weaker interactions are less symmetric.

Recent calculations~\cite{preprintVG:2016} demonstrate a considerable sensitivity of the 
high energy Compton scattering to the gravitational field's broken symmetries.
Using two Compton spectra measured by the HERA transverse polarimeter, a signature for 
gravitational ${\cal CP}$ violation has been extracted in the same ref.~\cite{preprintVG:2016}.
These spectra have been obtained from unpolarized laser--electron and 
left helicity laser--positron scatterings.
The reported upper limit of the gravitational 
parity ($\cal P$) violation is $(1.3\pm 0.3)\cdot 10^{-11}$ at 13~GeV energies.
At low energies, most of the existing ${\cal P}$-asymmetric gravity limitations are
model dependent and set by precise spectroscopic or polarized torsion 
pendulum~\cite{Heckel:2008hw} experiments.
They constrain hypotheses such as 
Lorentz violation~\cite{Kostelecky:2004pd}, torsion gravity~\cite{Hehl:1976kj}, 
exchange of pseudoscalar bosons~\cite{Moody:84}, and a few others.
A detailed review~\cite{Ni:2009fg} for low energy spin-dependent gravitation 
quotes a current best limit around  $10^{-7}$.
Stringent limits on space chirality, associated with parity violation, 
have been set by  astrophysical observations exploring linear polarization
of the light emitted from objects at cosmological distances~\cite{Gleiser:2001rm}.
The observational bounds limit the hypothetical dispersive vacuum birefringence down to
a level of $6\cdot10^{-37}$~\cite{Altschul:2011ab} or $5\cdot10^{-38}$~\cite{Stecker:2011ps} depending on the applied model or the source Gamma Ray Burst event.
These constrains, however, could hardly be applied to the gravitational fields with 
distinctive refractive (energy independent) nature.  

In this paper we will follow the formalism developed in ref.~\cite{preprintVG:2016}
to evaluate high energy laser-Compton scattering's analyzing power for 
gravity's left--right preference in a model-independent manner.
After a short description of the HERA transverse polarimeter
setup we will explore the polarized Compton spectra sampled during the 2004--2007 running period 
to extract the Compton edge left--right energy asymmetry and derive 
the magnitude of the gravitational space parity violation.
At the end we will discuss the detected ${\cal P}$-asymmetric gravitation's possible impact 
on some interpretations of astrophysical data.

\section{Gravitational  refractivity}

An external gravitational field with a Newtonian potential $U$ modifies the
momentum $P$ and energy $\cal E$ relation of a particle via the expression 
\begin{equation}
c\frac{P}{\cal E} = \frac{v}{c} - \frac{2 U}{c^2} + {\cal O}\Bigl(\frac{U^2}{c^4}\Bigr),
\label{refr}
\end{equation}
where $v$ is the speed of the particle and $c$ is the speed of light.
This gravitational refraction\footnote
{The terminology originates from optics (for $v=c$).
Since the right side of Eq.~(\ref{refr}) is energy independent, gravity is not dispersive.} 
is derived from GR, 
by combining Eq.(3) with Eq.(30) exact solution in 
ref.~\cite{Evans:2001hy},
and using the Newtonian potential \hbox{$U=-GM/R$}
for the particle at a distance $R$ from the gravitating mass $M$.
The combined result is approximated for weak fields using
expansion by powers of potential $U$ to obtain Eq.(\ref{refr}).   
\begin{table}[h]
\caption{
Gravitational fields and gradients (per one meter) at laboratory.
Listed are only the dominant contributors: the Earth, the Sun, the Milky Way, and the 
Local (Virgo) Supercluster.} 
\label{tab1}
\begin{ruledtabular}
\begin{tabular}{|c|c|c|c|c|}
  \diagbox{Potential}{Source} & Earth & Sun & Galaxy &  Virgo SC \\
\hline\hline
$U/c^2$ & $7\cdot10^{-10}$ & $9\cdot10^{-9}$ & $3\cdot10^{-7}$  & $3\cdot10^{-5}$\\
$\Delta U_R/c^2/m$ & $10^{-16}$& $7\cdot10^{-29}$ & $10^{-27}$ & $10^{-36}$\\
\end{tabular}
\end{ruledtabular}
\end{table}
Within GR all phenomena  are described by Einstein's equation, expressing
energy-momentum conservation in curved space-time.
In weak fields, Eq.~(\ref{refr}) together with energy-momentum conservation is 
sufficient for evaluating all gravitational effects.
In particular, for massless particles (light photon's gravitational deflection), 
relations similar to Eq.~(\ref{refr}) have been explored by many 
authors; see ref.~\cite{deFelice:1971ui} and references therein, 
or, for recent derivations, see refs.~\cite{Sen:2010zzf} and~\cite{Chu:2010tc}.
In a laboratory, the major attractors create fields with $U/c^2 \ll 1$
(see table~\ref{tab1}) and the resulting field $\Sigma U$ could also be  
described in terms of a flat space refractivity.
In Eq.~(\ref{refr}), the equivalence principle manifests itself by 
the independence of the gravitational constant $G$, and the potential in general,  
on whatever property of the particle.
This leads to cancellation of the gravitational potential 
in  energy-momentum conservation for any initial and final states at the quantum 
particles' interaction vertex.
Since any observable has to be gauge invariant, the  gravitational 
measurables can only depend on potential differences
\begin{equation}
\Delta U (G,M,R) = U\frac{\Delta G}{G}+ U\frac{\Delta M}{M} - U\frac{\Delta R}{R},
\label{du}
\end{equation}
where the first term $\Delta U_G\equiv U\Delta G/G$ violates  gravitational equivalence  
between different particles or states.
In GR $\Delta U_G = 0$ and experimentally it is constrained exceptionally by 
low energy tests.
The second term with mass change is currently not associated with any 
known test or system and the third term  $\Delta U_R \equiv U\Delta R/R $ is responsible 
for the conventional gravitational effects: the particles' deflection and 
frequency shift.
Typical magnitudes of these effects for relativistic particles  
at the Earth's surface 
are proportional to the potential difference $\Delta U_R$  presented in 
table~\ref{tab1}.
So far only the vertically moving (keV) photons' frequency change  
has been measured using the nuclear M\"ossbauer detectors~\cite{Pound:1960zz}.
High energy particles' (or light's) deflection \hbox{ $ 2L \Delta U_R/c^2  $}  
over a horizontal distance $L$ is out of reach of any laboratory instrumentation 
even for $L \sim$\ km scale.

In order to quantify and measure the space parity violation induced by gravitation, let's
assume a spin-dependent gravity with different couplings to the left and right helicity particles.
For this purpose, we introduce an interaction constant 
\begin{equation}
 G_s = G + \frac{\vec{s}\cdot\vec{P}}{P}\Delta G_{\cal P},
\label{sdg}
\end{equation}
which couples the gravitational field to the particles with spin $\vec{s}$.
This basic and minimal assumption will modify Eq.~(\ref{refr}) to
\begin{equation}
c\frac{P}{\cal E} =\frac{v}{c} -\frac{2}{c^2}\biggl(U+ \Delta U_s\biggr),
\label{refr2}
\end{equation}
with a space parity violating term
\begin{equation}
\Delta U_s = \frac{\vec{s}\cdot\vec{P}}{P}U\frac{\Delta G_{\cal P}}{G}
\label{dup}
\end{equation}
The helicity ($\vec{s}\cdot\vec{P}/{P}$) dependent interaction is assumed to be
small, so that $\Delta G_{\cal P} / G \ll 1$.
The specified mirror-symmetry breaking gravity will potentially affect all polarized 
interactions in a gravitational field through the relation~(\ref{refr2}), while
the second term in this relation is not universal anymore and depends on the
interacting particle's spin state.

\section{High energy Compton scattering in a gravitational field}

\begin{figure}[h]
\centering
\includegraphics[scale=0.47]{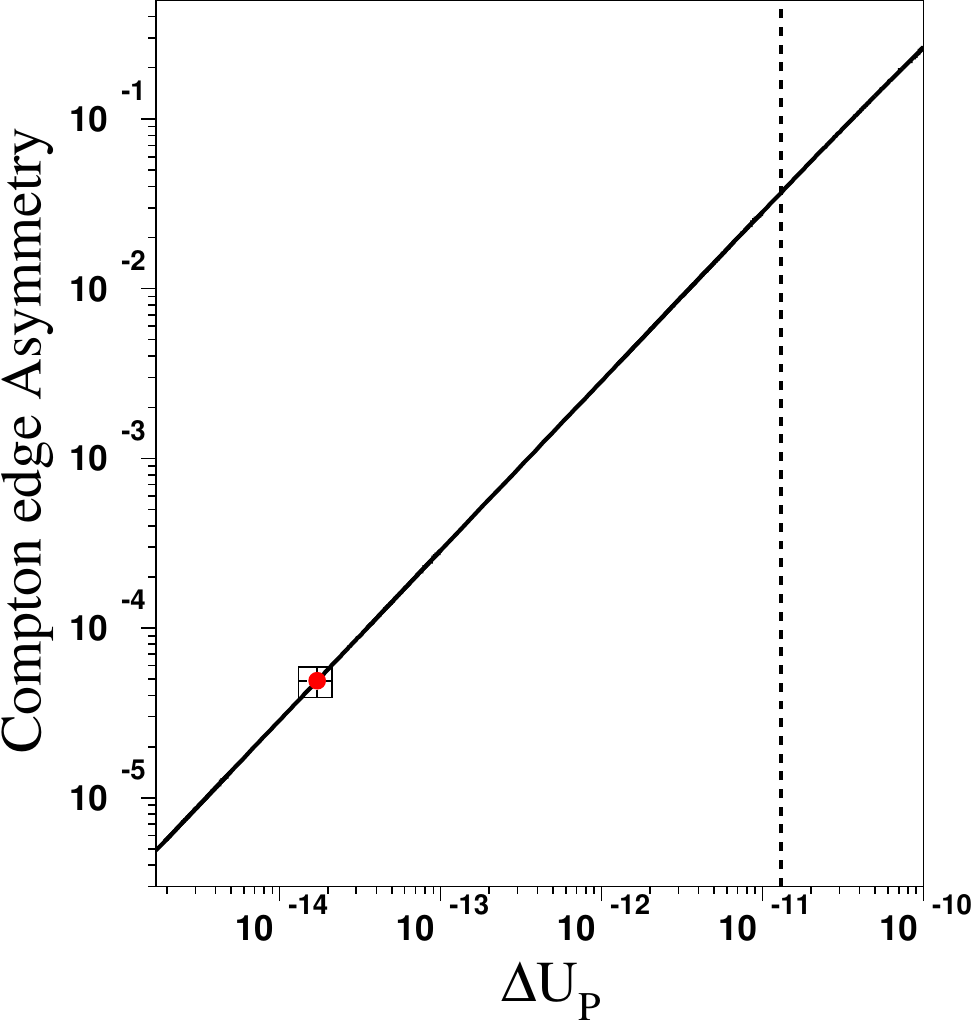}
\caption{\label{fig1}
The sensitivity of the Compton photons' maximum energy to the gravitational 
field helicity dependent coupling (potential difference 
\hbox{$\Delta U_{\cal P} $}) for the HERA 
transverse polarimeter.
The dashed vertical line indicates an upper limit detected  
in ref.~\cite{preprintVG:2016}.
The experimental result of this paper 
is shown by the point with error box.}
\end{figure}

Consider the polarized Compton process when a laser photon with 
energy $\omega_0$ and helicity $\lambda$ scatters off an accelerated lepton with 
high energy \hbox{$\cal E$, mass $m$ (Lorentz factor $\gamma \gg 1 $)}, and  zero helicity 
(from here on natural units are used).
Then, the scattered photon with maximum energy $\omega_{max}$ (at the Compton edge) 
will retain the initial 
helicity~~\cite{mcmaster:1961xe}
while the secondary lepton will acquire a helicity 
 \hbox{$\lambda_e =-\lambda x(4+2x)/(3x^2+6x+6)$}, according to 
refs.~\cite{Kotkin:2002ra, Lipps}\footnote
{A small change of lepton's helicity in the parity violating gravitational field is discussed below, in subsection \it{Systematic errors}.}.
The Compton kinematic factor \hbox{$x=4\gamma \omega_0\sin^2{(\theta_0/2)}/m$}, is defined
for an initial photon--lepton interaction angle $\theta_0$.
Assuming the scattering takes place in  
a gravitational field which  violates $\cal P$-parity by an amount of $\Delta U_s$,
we apply Eq.~(\ref{refr2}) to all initial and final particles.
After lengthy but simple calculations, the energy-momentum conservation 
reads 
\begin{equation}
x-y(1+x)-2(y^2 (\lambda_e-2) -2y(\lambda_e-1)+\lambda_e )\gamma^2\lambda  \Delta U_{\cal P} = 0,
\label{comp0}
\end{equation}
where $y=\omega_{max}/{\cal E}$ is the maximum relative energy of the scattered photon and
\hbox{$\Delta U_{\cal P}=U\Delta G_{\cal P}/G$}.
This relation with the lowest order non-vanishing gravitational term  
is valid for high energy Compton scattering down to the $\mathcal{O}(\gamma^{-4})$ 
terms (in the HERA experiment \hbox{$\gamma^{-4}=1.4\cdot10^{-19}$}). 
According to Eq.(\ref{comp0}), the maximum energy of the scattered gamma particle will depend on 
the laser helicity and the mirror symmetry breaking gravitation will induce a Compton edge   
asymmetry
\begin{equation}
A=\frac{\omega_{max}^--\omega_{max}^+}{\omega_{max}^-+\omega_{max}^+},
\label{asym1}
\end{equation}
where the upper indices denote the helicity states.
From Eq.~(\ref{comp0}) there follows
\begin{equation}
A=\frac{2u(x^2+2x+2)}{-x^4-4x^3-7x^2+(u-6)x+2u-2},
\label{asym2}
\end{equation}
with the assignment $u\equiv -2\gamma^2 \Delta U_{\cal P}$.
And, inversely, from a measured Compton edge spin asymmetry $A$ one can derive 
the gravitational left--right helicity potential's  difference
\begin{equation}
\Delta U_{\cal P} =\frac{1}{2\gamma^2}\cdot \frac{{A}(x^2+2x+2)(x+1)^2}{2x^2+ 
(4-{A})x-2{A}+4}.
\label{gpot}
\end{equation} 
In the above relations the amplification factor $\gamma^2$ sets the detection  
scale for the high energy Compton process to measure the gravitational parity violation.
In order to estimate the sensitivity of the laser-Compton scattering we insert 
the parameters of the HERA transverse polarimeter setup (see the next section) 
into Eq.~(\ref{asym2}).
For a range of feasible left--right helicity asymmetry measurements
one can refer to high energy precise detectors.
Measurements of asymmetries as low as $10^{-7}$ have been reported at 
SLAC 50~GeV~\cite{Anthony:2003ub,Anthony:2005pm} or at 
MAMI 1~GeV experiments~\cite{Maas:2005fk,Maas:2008zzc}.
Hence, for the 
sensitivity plot presented by Fig.~\ref{fig1}, we conservatively use a lower asymmetry 
value of $5\cdot 10^{-6}$.

\section{Experimental setup}

\begin{center}
\begin{figure*}[!ht]
\includegraphics[scale=0.7]{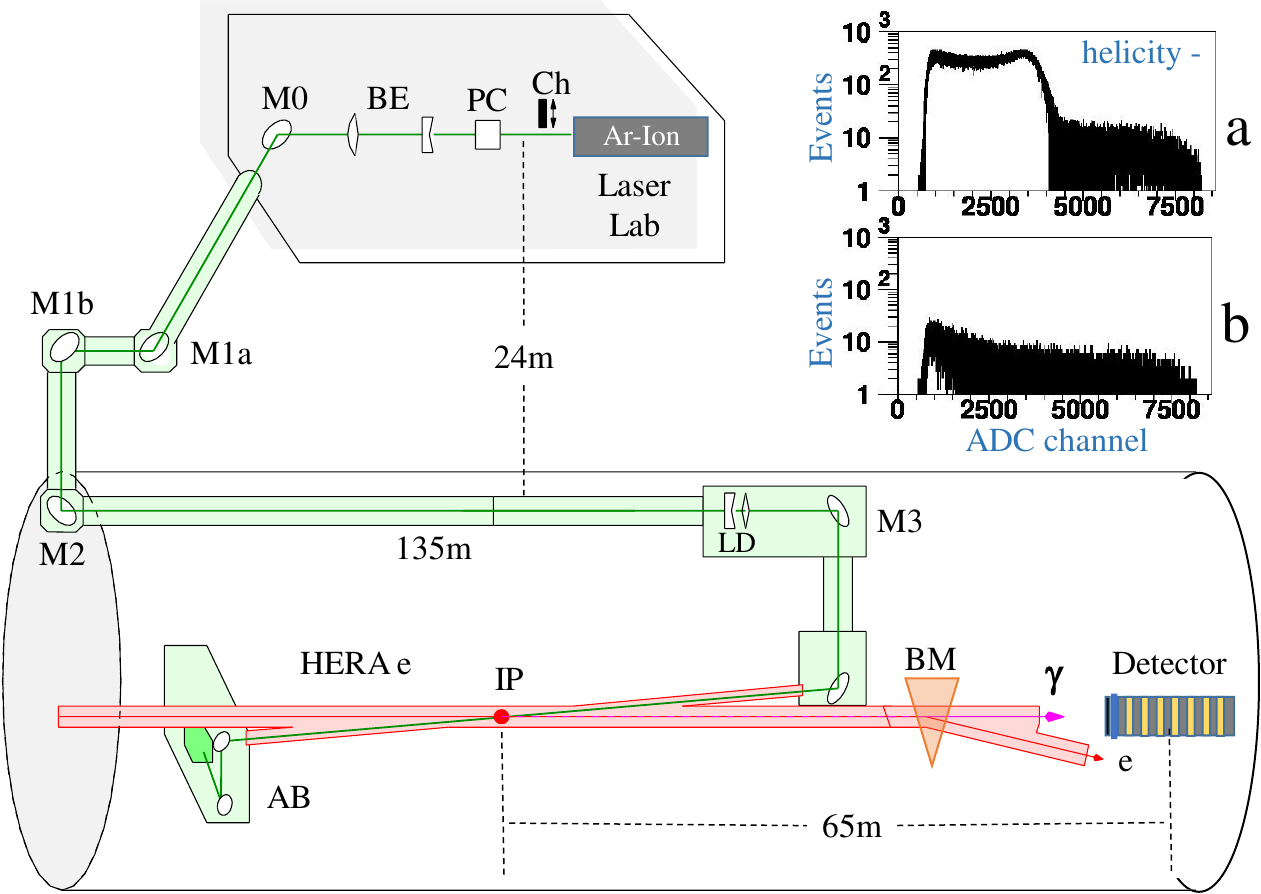}
\caption{\label{fig2}
Simplified outline of the HERA transverse polarimeter.
The setup
elements displayed in the Laser Lab and HERA tunnel are: light Chopper-shutter (Ch), 
Pockels Cell (PC), Beam Expander (BE), Mirrors (M0--M3),
Lens Doublet (LD), light Analyzer Box (AB), laser--lepton 
Interaction Point (IP) and Bending dipole Magnet (BM).
{\it Inset (a)}: Raw Compton spectrum measured with left helicity laser
during 45 sec shutter open cycle.
{\it Inset (b)}:  Background Bremsstrahlung spectrum 
sampled during 15 sec shutter closed period.}
\end{figure*}
\end{center}

The HERA transverse polarimeter is built to measure the average vertical 
spin of the circulating electrons or positrons using spatial and energy spectra 
from polarized laser Compton scattering.
The spectra have been sampled by directing 514.5~nm laser 
light against the HERA 27.6~GeV electron 
beam with a vertical crossing angle of 3.1~mrad and detecting the produced 
high energy $\gamma$-quanta with a segmented calorimeter.
The listed initial conditions correspond to the kinematic parameter $x=1.02$
and Compton edge $\omega_{max}=13.9$~GeV for the scattered photon beam.
A reduced layout of the 
polarimeter setup is shown in Fig.~\ref{fig2}.
The whole detection scheme is designed for the measurement of an up--down spatial
asymmetry of the $\gamma$-quanta which is introduced by a flip of the laser 
light's helicity and is proportional to the lepton beam transverse polarization.
The laser (Coherent Sabre Argon Ion) produces 10 W CW linearly polarized light
in $TEM_{00}$  mode.
The linear (linearly polarized) light is converted to circular (circularly polarized)  
by a Pockels 
cell with 85 Hz switching between positive and negative helicities.
A mechanical chopper periodically blocks the laser beam
shutting the light off for 15 sec within each 1 min measurement cycle   
in order to sample the background spectra.
An evacuated transport system with remotely controlled mirrors delivers the 
laser light about 200 m from the optical lab to the Compton interaction point (IP).
The laser beam is expanded (1:10) before the transport for focusing into the 
lepton beam by a lens-doublet installed 18.4 m upstream the IP.
 An analyzer optical 
setup at the laser beam-dump monitors the remnant of linear light to optimize
the circular polarization magnitude at the laser--lepton interaction point.

The energy measurement of the Compton $\gamma$-quanta
is auxiliary and serves as a means to enhance the spatial asymmetry by imposed
energy cuts.
Since we are going to compare the maximum energies of the photons from the Compton spectra
tagged by light helicity we 
concentrate on those details of the experimental setup that are important for energy
measurement only, ignoring all features related to the lepton polarization detection.

The scattered Compton photons originate from an interaction region (IR) 
about $0.5$~m long, defined by the crossing angle and size of the electron and 
laser beams.
Bending dipole magnets downstream of the IR separate the electron 
and $\gamma$ beams and the photons leave the vacuum pipe through a 
0.5~mm thick aluminum window to travel via a mostly evacuated 39~m path  
before entering the 
calorimeter, which is installed 65~m downstream  the IR.
Collimators placed at a distance of 47~m from the IR define an aperture of 
$\pm 0.37$~mrad, the same as the angular size of the calorimeter as seen from 
the IR.
The aperture is 15 times larger than the largest (horizontal) angular 
spread of electrons at the IR and 40 times larger than the characteristic 
radiation angle $1/\gamma$, so the acceptance inefficiency can be ignored.
The collimators are followed by magnets to sweep out any charged background.
    
The calorimeter consists of 12 layers of 6.2-mm thick tungsten and 2.6-mm
thick scintillator plates surrounded by four wavelength shifters attached to four
photomultipliers (PMT).
A converter-preshower tungsten plate with one radiation length, 
installed 
in front of the calorimeter, provides charged particles for the operation of 
position-sensitive 
silicon strip detectors.
 
PMT signals from single photons are stretched by shapers to 96~ns and fed into 
analog-to-digital converter (ADC) in a 10 MHz data acquisition (DAQ) 
system similar to the HERA cavity 
polarimeter DAQ described in ref.~\cite{Baudrand:2010hp}.
An essential difference from the cavity polarimeter DAQ is the trigger mode 
operation, which stops the ADC pipeline only for the signals exceeding a given 
threshold (about 3~GeV).

The detector performance has been simulated with the GEANT Monte Carlo program and 
tested using DESY and CERN test beams.
The measured energy resolution of 
24\%~GeV$^{1/2}$, spatial non-uniformity of $\pm 1\%$, and nonlinearity of $2\%$ at 
20~GeV are  in agreement with the simulations.

Apart from the laser light, the lepton beam also interacts with residual gas,
thermal photons, and the bending magnetic field in the beam pipe, producing, respectively,
Bremsstrahlung, scattered blackbody radiation, and synchrotron radiation reaching the 
calorimeter.
To measure this background, the laser beam is blocked for 15~sec
of each 1~min measurement cycle (laser light on/off is 45/15~sec).
This procedure 
allows eliminating the background by a simple subtraction of time normalized 
light-off spectrum from the light-on spectrum.
The exact on/off durations are 
counted by DAQ clocks.

 At the time of the measurements, the average Compton $\gamma$ rate  was 37.6~kHz
above the energy threshold of 3~GeV, while the background rate was
3.6~kHz.
The rate distributions obey Landau rather than Gauss statistics, with 
longer tails towards lower rates.
With such a high threshold, only the Bremsstrahlung contributes to the background since 
the maximum energy of the scattered blackbody radiation is 0.73~GeV and the synchrotron
radiation is absorbed in the preshower and the first tungsten plate of the calorimeter.

Additional details about the setup are available in 
refs.~\cite{Barber:1992fc,Lomperski:1993aw,Gharibyan:2003fe}
\begin{figure}[h]
\centering
\includegraphics[scale=0.47]{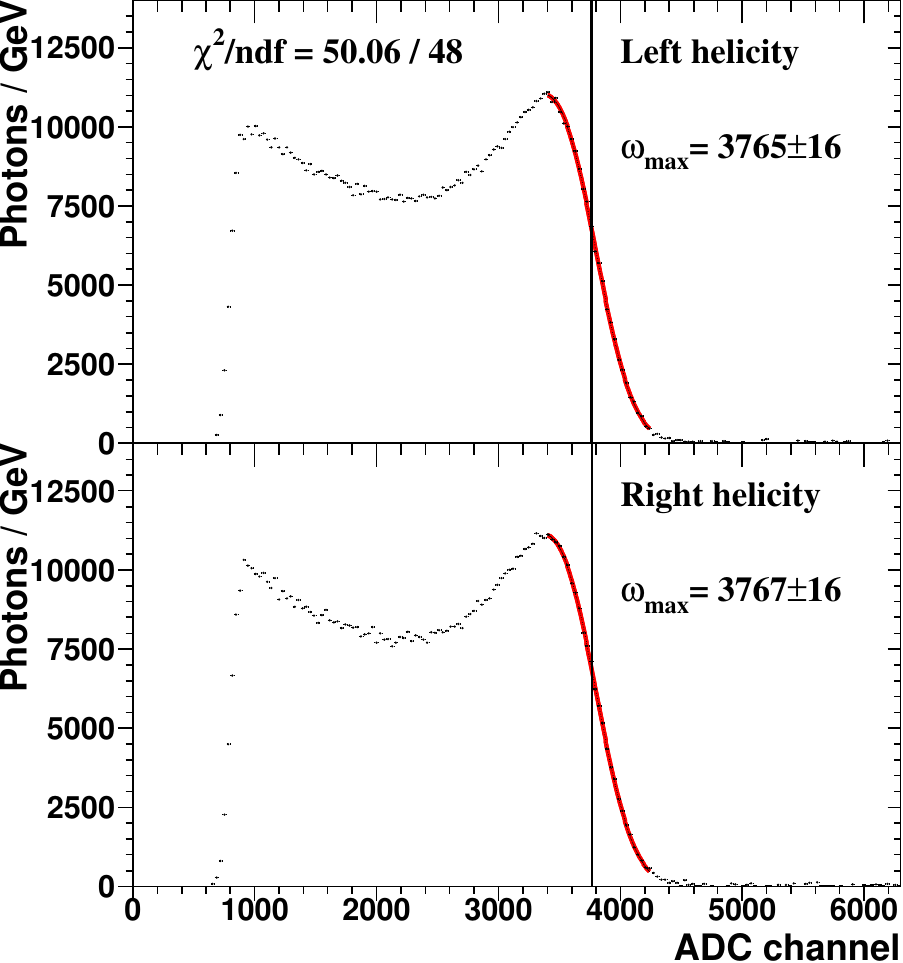}
\caption{\label{fig3}
Fitted Compton spectra from Fig.~\ref{fig2} {\it Inset}.}
\end{figure}

\section{Experimental Results}

{\it Method of Analysis.}
The raw spectra measured by the HERA transverse polarimeter are histograms 
with 8192 channels each (13 bit ADC).
An example of a per-minute spectrum measured at 
30/11/2006 during \hbox{23:41:33--23:42:33} is presented in Fig.~\ref{fig2} {\it Inset}.
For this example, the laser-on (off) rate was 50.1(1.6)~kHz and $1.1\cdot 10^{6}$
high energy photons were collected for each helicity state (the {\it Inset a} displays 
only the left helicity spectrum).
Algorithms for extracting the absolute maximum photon energy from the Compton and Bremsstrahlung 
spectra are described in ref.~\cite{preprintVG:2016,Gharibyan:2003fe}.
Here we are interested in the Compton edge left-right helicity asymmetry so, we apply 
a simplified analyzing algorithm involving only the Compton spectra.
After a reduction
of the background subtracted Compton spectra to 256 bins both left and right spectra
are fitted in a single \hbox{MINUIT}~\cite{James:1975dr} run.
The fitting function
for each spectrum is a convolution of the theoretical 
Compton energy distribution ${d\Sigma}/{d\omega}$ with the 
detector response Gaussian function
\begin{equation}
F(E_a)=N_\lambda \int^{\omega_{\lambda}}_{0} 
\frac{d\Sigma}{d\omega}\frac{1}{\sqrt{\omega}} 
\exp\Biggl({\frac{-(\omega-CE_a)^2}
{2\sigma_0^2 \omega}\Biggr) d\omega},
\label{eqfold}
\end{equation}
where 
$\sigma_0$  and $E_a$  denote the
calorimeter resolution and detected photon's energy in ADC units respectively.
The normalizing factors $N_{+,-}$ and the maximum Compton energies $\omega_{+,-}$ are 
spectrum dependent.
 These four variables together with $\sigma_0$ and the calibration 
factor $C$ are free parameters of the fit.
The last two (detector) parameters can (slightly) 
change in the course of the lepton-fill or from fill to fill.
The integral (\ref{eqfold}) for each \hbox{MINUIT} cycle
is calculated numerically using Gaussian quadrature.
The Compton edge extraction fitting procedure is applied only to data
within a narrow energy slice around the maximum energy to avoid contamination 
from systematic effects in the lower energy bins (see ref.~\cite{Gharibyan:2003fe}).
Examples of fitted spectra (the same data as in Fig.~\ref{fig2} {\it Inset}) together with 
the fit outcome are displayed in Fig.~\ref{fig3}.
The quoted uncertainties are the statistical errors calculated by the fitting 
routine.
\begin{figure}[h]
\centering
\includegraphics[scale=0.58]{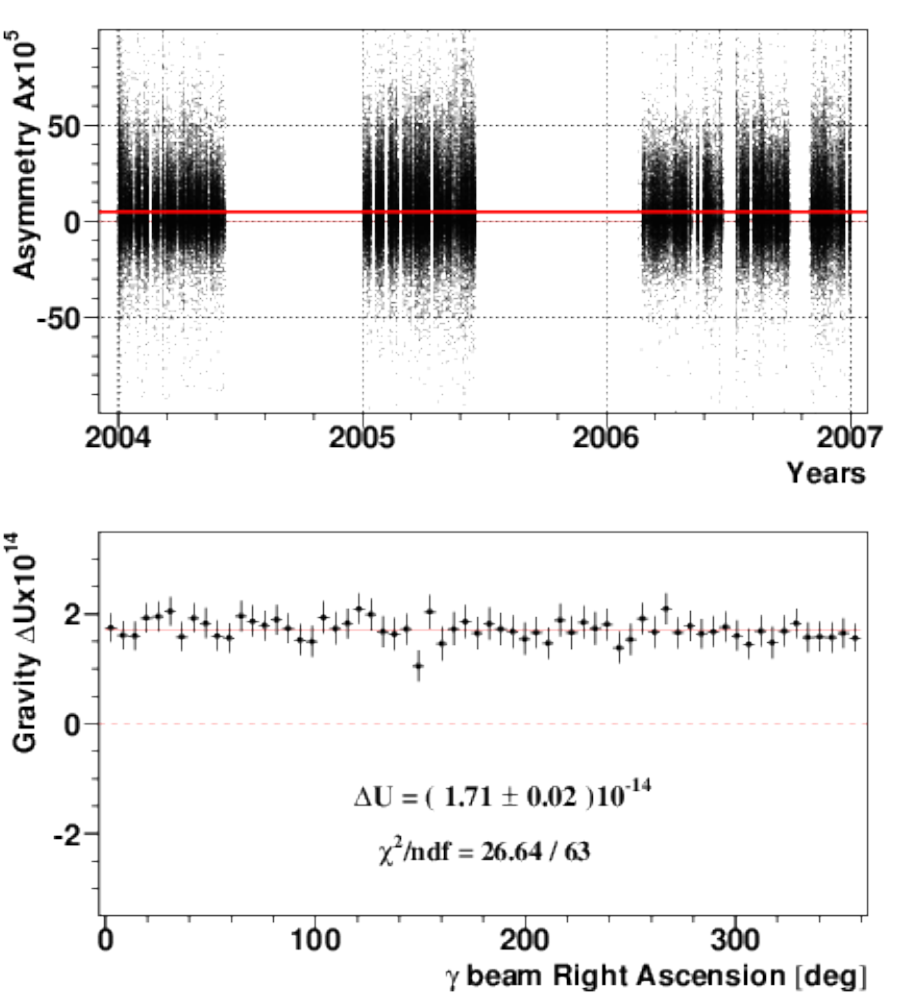}
\caption{\label{fig4}
{\it Upper plot:} Compton edge helicity dependent asymmetry measured by HERA transverse polarimeter.
{\it Lower plot:} Angular dependence of the observed
gravitational left--right potential difference.}
\end{figure}

{\it Outcome of the measurements.}
Applying the described Compton edge extraction procedure to
data collected during the 2004--2007 HERA running period, we
succeeded in fitting 314,896 spectra complying  with the standard
criterium for $p$-values to exceed the threshold of 5\%.
Combining left and right helicity Compton edge values for each minute 
according to Eq.~(\ref{asym1})
we obtained 157,448 asymmetry points, plotted in Fig.~\ref{fig4}
against their measurement time.
Two extended gaps between the data points
correspond to the accelerator shutdown periods.
The average magnitude of the measured asymmetries
weighted by  inverse statistical errors is 
\hbox{  $(4.89 \pm 0.10 )\cdot 10^{-5}$}.
From the observed asymmetry, we  derive  the  gravitational 
left--right potentials difference $\Delta U_{\cal P}$ via Eq.~(\ref{gpot}).
We also  explore per-minute timestamps of the measurements to check how 
the asymmetry, or  values of $\Delta U_{\cal P}$, are grouped relative to a 
fixed direction in space.
For this purpose, we convert  the timestamps $T$ to 
Right Ascension (RA) angles in the celestial coordinate system by the formula
\hbox{$RA= 360^\circ \cdot T\pmod {T_S} -  9.87^\circ$}, where $T_S$ is 
the duration of the sidereal day and the angular offset is the polarimeter's
setup longitude.
While the rotation of the accelerator with the Earth sweeps the Compton beam along a 
circle on the celestial sphere, the declination angle of $33.35^\circ$ stays 
constant.
The resulting angular dependence reduced to 64 bins is shown 
in Fig.~\ref{fig4}.

{\it Systematic errors.}
The directional independence of the measured asymmetry is a validity check for
the applied model.
Indeed, observation of any preferred direction in space would 
violate the Lorentz symmetry (SR).
Describing such a vectorial asymmetry that breaks 
the equivalence principle of GR together with the Lorentz invariance of SR  would need a more
complicated formalism than the applied scalar gravitational potentials' formulas.
As possible candidates for vectorial theories one can consider the Standard Model Extension
(SME) with post-Newtonian approximation~\cite{Kostelecky:2003fs,Bailey:2006fd} or the  
Chern--Simons General Relativity Extension~\cite{Jackiw:2003pm,Smith:2007jm}. 
The observed asymmetry, however, is highly isotropic and this type of theoretical 
errors can be  safely ignored.
We came to this conclusion by fitting the observed 
angular dependence with a constant for different binnings and 
comparing $\chi^2$ to ndf 
(the values for 64 bins are displayed in Fig.~\ref{fig4}).
A similar insensitivity to the celestial angular coordinates have been observed in the ESRF
laser--Compton (unpolarized) kinematic edge distribution~\cite{Bocquet:2010ke}, 
which limits few parameters of the space isotropy violation in SME~\cite{Colladay:1998fq}.

The conventional gravity spin effects introduced by Earth's rotation are
described in refs.~\cite{Wald:1972sz,Papapetrou:1951pa,Bonnor:2002zg}
for classical or macroscopic test particles.
For elementary particles, an estimate of the additional 
energy induced by Earth's spin is 
\hbox{$\hbar g/c=2.2\cdot 10^{-23}$eV~\cite{Obukhov:2000ih}}, 
where $g$ is the gravitational acceleration in the laboratory.
Compared to the measured asymmetry, this contribution  is vanishingly small.
    
Besides gravitation, other interactions also may alter 
the energy--momentum relation~\cite{Latorre:1995cv,Dittrich:1998fy}.
For high energy processes, the main competitor to gravity would be a 
background electromagnetic field.
As an example, let's estimate the influence of magnetic fields.
According to Eq.~(1.1) from ref.~\cite{Latorre:1995cv}, 
the maximum impact on the energy--momentum relation for photons 
in a magnetic field $B$ is given by
\begin{equation}
\frac{P}{\cal E} = 1 - \frac{11}{45} \frac{\alpha^2}{m^4} B^2 ,
\label{emref}
\end{equation} 
\noindent 
where $\alpha$ is the fine structure constant.
Hence, the refractivity created by a 4~T superconducting magnet
is about $8\cdot 10^{-21}$ and that of Earth's  
magnetic field is $10^{-25}$.
Magnetic refractivity's (transverse) spin 
asymmetry amounts to about one-half of the mentioned values or, more precisely, 
to the 6/11-th part.
These tiny effects are still experimentally unreachable, and, compared to 
the magnitude of the gravitational potential 
(either the Local Supercluster's $\sim 10^{-5}$ or the Earth's $\sim 10^{-9}$) 
in Eq.~(\ref{refr}), are completely negligible.
The hypothetical influence of the helicity-dependent weak interaction is excluded
by energy conservation: the weak bosons are too heavy to contribute 
to the Compton scattering at the HERA lepton energy scale,
and, no virtual tree or loop heavy boson can shift the Compton edge.
Thus, an impact on the Compton edge asymmetry from non-gravitational forces is largely ignorable.
\begin{figure}[!ht]
\centering
\includegraphics[scale=0.55]{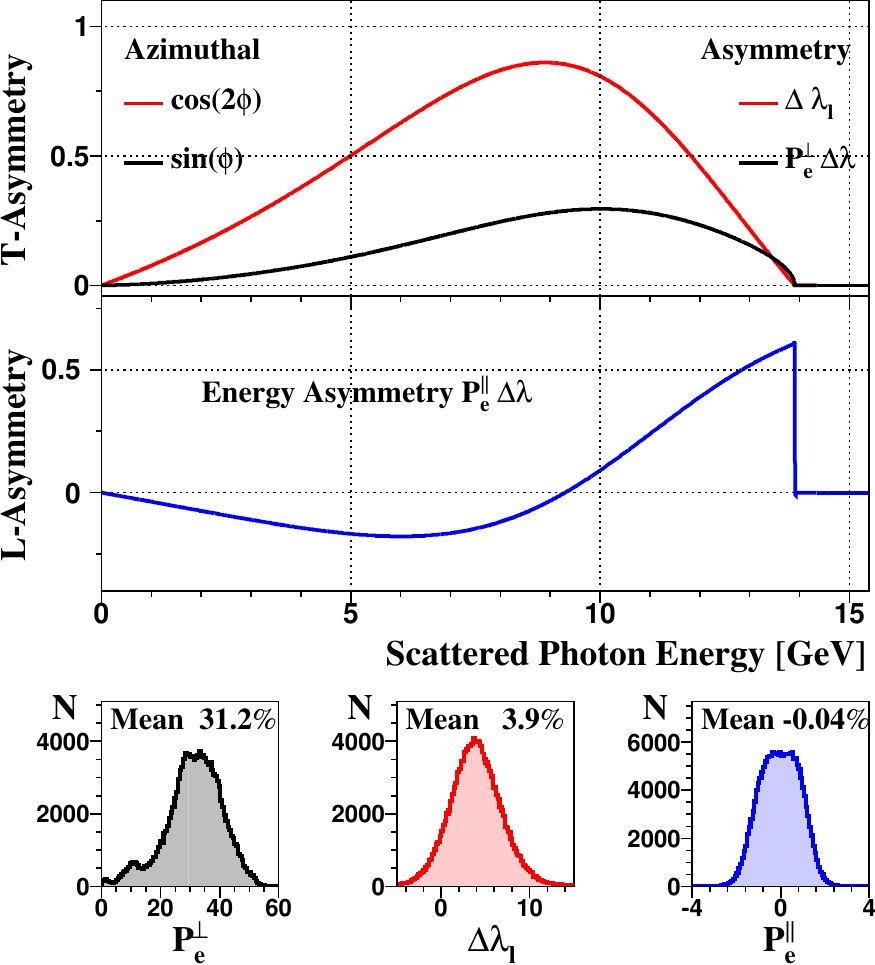}
\caption{\label{fig5}
{\it Upper rows:} Transverse-spatial (angular) and
longitudinal-energy  asymmetries in Compton scattering 
associated with the initial photon's polarization (linear $\lambda_l$ and circular $\lambda$) flips.
Analyzing powers (asymmetries for the cases 
\hbox{$
\lvert P_e^\parallel \Delta\lambda\rvert=1,
\lvert \Delta\lambda_l\rvert=1,
\lvert P_e^\perp \Delta\lambda\rvert=1
$}) of the 
HERA transverse polarimeter setup.
{\it Lower row:} Percent polarizations  measured during 2004--2007.
}
\end{figure}

Possible instrumental false asymmetries may arise from polarized effects 
correlated with the laser helicity flips.
These are linked with the lepton's  transverse or longitudinal spin 
and the laser light's linearly polarized fraction coupled to the gamma-calorimeter segmentation.
The maximum possible theoretical asymmetries in polarized Compton scattering at the HERA polarimeter 
are presented in Fig.~\ref{fig5}.
As follows from the plots, at the Compton edge energy, the
influence of the longitudinal polarization of the electron beam ($P_e^\parallel$) is dominant while the 
contributions of the transverse polarization ($P_e^\perp$) and linear light
($\lambda_l$) drop to zero.
Although none of the mentioned factors could physically alter the Compton gamma's maximum energy, the 
instrumental smearing can mimic 
a Compton edge shift, given a sufficiently 
large lepton beam polarization or linear light.
In order to estimate the magnitude 
of these effects, we explored polarimeter detector simulation codes used in 
ref.~\cite{Gharibyan:2006note} with  $P_e^\perp$, $\Delta\lambda_l$ and 
$P_e^\parallel$ measurements (Fig.~\ref{fig5} {\it Lower row}).
From simulated per-minute spectra ($1.2\cdot 10^6$ photons per helicity state), the Compton edge asymmetry 
has been extracted applying the same algorithm as for the real data.
For the simulated 100\% 
 $P_e^\perp$, $\Delta\lambda_l$ and $P_e^\parallel$ values
 the instrumental false asymmetries amount to  
$1.013\cdot 10^{-5}$, $5.379\cdot 10^{-6}$ and $4.625\cdot 10^{-3}$ respectively.
 These numbers have been 
derived from 60,000 simulated spectra for each factor.
Scaling the asymmetries by the observed average values for the 2004--2007 running period
$< P_e^\perp >=31.18\%$,  $< \Delta\lambda_l > =3.92\%$ and 
$<P_e^\parallel >=-0.038\%$, we get an estimate of the measured asymmetry 
instrumental error of $3.62\cdot 10^{-6}$, where we chose a quadratic summation to neutralize the 
negative sign of the mean longitudinal polarization.
Alternatively one can correct the observed asymmetry by the longitudinal 
polarization factor, assigning half of the correction magnitude as systematic error,  
and apply more conservative linear summation.
This will enhance the observed asymmetry from $4.89\cdot10^{-5}$ to
$5.07\cdot10^{-5}$  and will rise the instrumental error from  
$3.62\cdot 10^{-6}$ to $4.25\cdot 10^{-6}$.
The most conservative method, 
however, is a linear summation of the errors' absolute values
which brings the maximum instrumental error to $5.13\cdot 10^{-6}$.
Using this value together with the above quoted  statistical uncertainty 
we get the measured asymmetry with an overall error
of \hbox{$(4.89\pm 0.52)\cdot10^{-5}$}.

In order to evaluate the uncertainty of the $\Delta U_{\cal P}$ one needs 
to propagate the measurement errors from the asymmetry as well the kinematic and 
Lorentz factors through Eq.~(\ref{gpot}).
Listing the errors of the  
$x$ and $\gamma$ factors' constituents yields 
$\sigma (\omega_0)/\omega_0\approx 10^{-5}$,
$\sigma (m)/m \approx 3\cdot10^{-7}$,
$\Delta (\theta_0)\!\approx\!2~mrad\Rightarrow$ 
$\Delta\sin^2{(\theta_0/2)}\approx 3\cdot10^{-6}$,
$\sigma ({\cal E})/{\cal E}\approx 10^{-3}$,
and we note that the dominant uncertainty comes from 
the HERA leptons' energy spread.
We have also estimated the contribution from the final state lepton polarization change 
$$\frac{\lambda_e - \lambda_e^{(u)}}{\lambda}=
\frac{8x(x^3+2x^2+2x+2)}{3(1+x)(x^2+2x+2)^2}u \approx 3.7\cdot 10^{-5},  $$
in the gravitational field, using the measured value of $u$. 
The scattered lepton polarization in parity violating gravitational field, $\lambda_e^{(u)}$,
is obtained using the modified kinematic relation~(\ref{comp0}) in the conventional sums over 
the lepton spin states 
\begin{align*}
 \lambda_e^{(u)} = \frac{\sum_{\lambda_e}d\sigma_y(\lambda,\lambda_e)\lambda_e}
{\sum_{\lambda_e}d\sigma_y(\lambda,\lambda_e)}, 
\end{align*}
where the $d\sigma_y(\lambda,\lambda_e)$ is the polarized Compton (differential)
cross-section from ref.~\cite{Kotkin:2002ra}. Thus, gravitational parity violation 
contributes a relative $9.2\cdot 10^{-5}$ enhancement to the scattered lepton's 
$40.374\%$ polarization.

The quoted systematic errors are collected in   table~\ref{tab2}.
\begin{table}[htb]
\caption{Error bank of the measurements of the Compton edge asymmetry and difference between the gravitational potentials.} 
\label{tab2}
\begin{center}
\begin{tabular}{|l|c|c|}
\hline
Source                           & Magnitude  & Error $\Delta A$   \\
\hline\hline
Magnetic field & $50~\mu T$ & $5.5\cdot 10^{-26}$ \\
Earth's rotation influence & $2\cdot 10^{-23}~eV$ & $1.6\cdot 10^{-29}$ \\
Transverse polarization & 0.31 & $3.2\cdot 10^{-6}$ \\
Laser linear polarization & 0.039 & $2.1\cdot 10^{-7}$ \\
Longitudinal polarization & -0.0004 & $1.8\cdot 10^{-6}$ \\
\hline
                  & Magnitude (rel.) & Error of $\Delta U_{\cal P}$   \\
\hline
Statistical fluctuations & $2.1\cdot 10^{-2}$ & $3.6\cdot 10^{-16}$ \\
Laser frequency shift& $10^{-5}$   &$1.7\cdot 10^{-19}$\\
Electron mass error & $3\cdot10^{-7}$ &$1.0\cdot 10^{-20}$\\
Interaction angle drift & $3\cdot10^{-6}$ &$5.2\cdot 10^{-20}$\\
HERA-e energy spread& $10^{-3}$ & $3.8\cdot 10^{-17}$ \\
e' spin correction& $9.2\cdot 10^{-5}$ & $1.5\cdot 10^{-19}$ \\
Asymmetry measurement & 0.11 & $1.9\cdot 10^{-15}$ \\
\hline
\end{tabular}
\end{center}
\end{table}
Calculations with the displayed values give the measurement's final result:
\hbox{$\Delta U_{\cal P}=(1.71\pm 0.036 \pm 0.187  )\cdot10^{-14}$},
where the statistical and systematic errors are displayed separately.
Substituting the obtained value in Eq.(\ref{dup}), for sum of potentials from
table~\ref{tab1}, we find       
$$\Delta G_{\cal P}/G=(5.7\pm 0.6 )\cdot10^{-10}$$.

This measurement, as a by-product of the HERA polarimetry, may also suffer from 
hidden systematic factors such as the major suspects -- the detector's helicity 
dependent gain or the lepton beam non-gaussian deviations in 6D phase-space 
convoluted with a possible laser spatial jitter at IP which is correlated 
with the light polarization~\cite{priv:Brinkmann}.
For the gain effects we have estimates from the bremsstrahlung edge 
fits~\cite{Gharibyan:2003fe} to be  
of the opposite sign (on average) to the observed asymmetry. Detailed corrections 
which on average will enhance the observed asymmetry, are, however, complicated 
since the bremsstrahlung and Compton beams have an unknown mutual offset on the 
face of the spatially inhomogeneous calorimeter.
For the lepton and the polarized laser beam spatial convolution the estimations are 
more complex if the optical jitter helicity dependent magnitude,  
averaged over the four years period, would differ considerably from a plausible 
zero value.
Such a deviation would, however, affect the spatial helicity 
asymmetry used for the transverse polarization measurement,
and is limited by the zero polarization measurements~\cite{Gharibyan:2006note}.

\section{Chiral gravity implications}

The main message of the observed parity violation is that 
gravitational equivalence is broken and systems or processes involving spin
are gravity dependent.
Although experimental detection is made at 
high energy, 13.9~GeV, gravity's universality, infinite range, and all-attractive nature
suggest an energy independence of the observed helicity preference of gravitation.

This would mean that the hydrogen atom's well known electron--proton spin 
radiation (21~cm line)  will be affected by the gravitational 
potential at the location of the atom.
So far, the observed deviations from the 21~cm line have 
been attributed to Doppler shifted velocities to infer 
the presence of dark matter~\cite{Persic:1995ru}.
Meanwhile, the detected chiral gravity could possibly explain the galaxies' anomalous 
rotational curves by differences in the gravitational potentials at the 
Earth (where the 21~cm line is calibrated) and the periphery of the galaxy.
Detailed calculations are outside the scope of this paper.
Besides, 
the uncertainties in the gravitational potentials, whether in galaxies or 
even at the Earth, seem too large for an accurate answer.

At scales larger than the atomic, one could possibly find an influence of 
chiral gravity on asymmetric chemical and helical molecules of life.
These, however, are composite complex systems and the interaction is more 
difficult to justify and quantify.

\section{Conclusions}

The indication of a spin dependence of gravitation, found earlier in a limited
amount of data, has been confirmed by the analysis of four years of data from the HERA transverse polarimeter.
A left--right helicity asymmetry at the kinematic edge
of the Compton scattered $\gamma$-quanta is established with a $9\sigma$
confidence level.
According to the described formalism, this asymmetry is induced by the breaking of
gravitational equivalence: the interaction's intensity depends on the helicity.
The contributions of the other three fundamental interactions to this asymmetry 
has been proven to be negligibly small.

Further experimental investigations of gravity's spin preference could 
be done at high-$\gamma$ accelerators or with atomic and nuclear
quantum experiments if sufficient sensitivity could be achieved.
The latter possibility assumes an energy independence of the observed effect,
which would suggest revisiting the hydrogen 21~cm line spectroscopy data
so as to disentangle the contributions of the gravitational fields from that of the Doppler velocities.

\section*{Acknowledgement}
We are thankful to the machine staff and all H1, \hbox{HERMES}, ZEUS collaborators who have contributed
to the setup and data taking of the HERA transverse polarimeter.

\end{document}

%% file: cauthors.tex
% Institute Addresses 

\def\greins{\affiliation{I. Physikalisches Institut der RWTH, Aachen, Germany }} 
\def\grzwei{\affiliation{ Institute of Physics and Technology of Ministry of Education and Science of Kazakhstan, Almaty, \mbox{Kazakhstan} }} 
\def\grdrei{\affiliation{ NIKHEF and University of Amsterdam, Amsterdam, Netherlands }} 
\def\grvier{\affiliation{ Argonne National Laboratory, Argonne, Illinois 60439-4815, USA }} 
\def\grfunf{\affiliation{Vinca Institute of Nuclear Sciences, Belgrade, Serbia }} 
\def\grsechs{\affiliation{ Andrews University, Berrien Springs, Michigan 49104-0380, USA }} 
\def\grsieben{\affiliation{School of Physics and Astronomy, University of Birmingham, Birmingham, United Kingdom }} 
\def\gracht{\affiliation{ INFN Bologna, Bologna, Italy }} 
\def\grneun{\affiliation{ University and INFN Bologna, Bologna, Italy }} 
\def\grzehn{\affiliation{ Physikalisches Institut der Universit\"at Bonn, Bonn, Germany }} 
\def\grelf{\affiliation{ H.H.~Wills Physics Laboratory, University of Bristol, Bristol, United Kingdom }} 
\def\grzwölf{\affiliation{ Institute for High Energies ULB-VUB, Brussels; Universiteit Antwerpen, Antwerpen; Belgium }} 
\def\grdreizehn{\affiliation{National Institute for Physics and Nuclear Engineering (NIPNE), Bucharest, Romania }} 
\def\grvierzehn{\affiliation{ Panjab University, Department of Physics, Chandigarh, India }} 
\def\grfunfzehn{\affiliation{ Department of Engineering in Management and Finance, Univ. of the Aegean, Chios, Greece }} 
\def\grsechzehn{\affiliation{ Physics Department, Ohio State University, Columbus, Ohio 43210, USA }} 
\def\grsiebzehn{\affiliation{ Calabria University, Physics Department and INFN, Cosenza, Italy }} 
\def\grachtzehn{\affiliation{ The Henryk Niewodniczanski Institute of Nuclear Physics, Polish Academy of Sciences, Cracow, }} 
\def\grneunzehn{\affiliation{ Faculty of Physics and Applied Computer Science, AGH-University of Science and \mbox{Technology}, Cracow, Poland }} 
\def\grzwanzig{\affiliation{ Department of Physics, Jagellonian University, Cracow, Poland }} 
\def\greinzwanzig{\affiliation{ Kyungpook National University, Center for High Energy Physics, Daegu, South Korea }} 
\def\grzweizwanzig{\affiliation{Rutherford Appleton Laboratory, Chilton, Didcot, United Kingdom }} 
\def\grdreizwanzig{\affiliation{ Institut f\"ur Physik, TU Dortmund, Dortmund, Germany }} 
\def\grvierzwanzig{\affiliation{Joint Institute for Nuclear Research, Dubna, Russia }} 
\def\grfunfzwanzig{\affiliation{ INFN Florence, Florence, Italy }} 
\def\grsechszwanzig{\affiliation{ University and INFN Florence, Florence, Italy }} 
\def\grsiebenzwanzig{\affiliation{ Fakult\"at f\"ur Physik der Universit\"at Freiburg i.Br., Freiburg i.Br., Germany }} 
\def\grachtzwanzig{\affiliation{CEA, DSM/Irfu, CE-Saclay, Gif-sur-Yvette, France }} 
\def\grneunzwanzig{\affiliation{ Department of Physics and Astronomy, University of Glasgow, Glasgow, United \mbox{Kingdom} }} 
\def\grdreisig{\affiliation{ Institut f\"ur Experimentalphysik, Universit\"at Hamburg, Hamburg, Germany }} 
\def\greindreisig{\affiliation{ Deutsches Elektronen-Synchrotron DESY, Hamburg, Germany }} 
\def\grzweidreisig{\affiliation{ Max-Planck-Institut f\"ur Kernphysik, Heidelberg, Germany }} 
\def\grdreidreisig{\affiliation{Physikalisches Institut, Universit\"at Heidelberg, Heidelberg, Germany }} 
\def\grvierdreisig{\affiliation{ Kirchhoff-Institut f\"ur Physik, Universit\"at Heidelberg, Heidelberg, Germany }} 
\def\grfunfdreisig{\affiliation{ Nevis Laboratories, Columbia University, Irvington on Hudson, New York 10027, USA }} 
\def\grsechsdreisig{\affiliation{ Institute for Nuclear Research, National Academy of Sciences, and Kiev National University, Kiev, Ukraine }} 
\def\grsiebendreisig{\affiliation{ Institute of Experimental Physics, Slovak Academy of Sciences, Ko\v{s}ice, Slovak Republic }} 
\def\grachtdreisig{\affiliation{ Jabatan Fizik, Universiti Malaya, 50603 Kuala Lumpur, Malaysia }} 
\def\grneundreisig{\affiliation{ Chonnam National University, Kwangju, South Korea }} 
\def\grvierzig{\affiliation{Department of Physics, University of Lancaster, Lancaster, United Kingdom }} 
\def\greinvierzig{\affiliation{Department of Physics, University of Liverpool, Liverpool, United Kingdom }} 
\def\grzweivierzig{\affiliation{ Physics and Astronomy Department, University College London, London, United \mbox{Kingdom} }} 
\def\grdreivierzig{\affiliation{Queen Mary and Westfield College, London, United Kingdom }} 
\def\grviervierzig{\affiliation{ Imperial College London, High Energy Nuclear Physics Group, London, United \mbox{Kingdom} }} 
\def\grfunfvierzig{\affiliation{ Institut de Physique Nucl\'{e}aire, Universit\'e Catholique de Louvain, Louvain-la-Neuve, \mbox{Belgium} }} 
\def\grsechsvierzig{\affiliation{Physics Department, University of Lund, Lund, Sweden }} 
\def\grsiebenvierzig{\affiliation{Departamento de Fisica Aplicada, CINVESTAV, M\'erida, Yucat\'an, Mexico }} 
\def\grachtvierzig{\affiliation{Departamento de Fisica, CINVESTAV, M\'exico City, Mexico }} 
\def\grneunvierzig{\affiliation{ Department of Physics, University of Wisconsin, Madison, Wisconsin 53706, USA }} 
\def\grfunfzig{\affiliation{ Departamento de F\'{\i}sica Te\'orica, Universidad Aut\'onoma de Madrid, Madrid, Spain }} 
\def\greinfunfzig{\affiliation{CPPM, CNRS/IN2P3 - Univ. Mediterranee, Marseille, France }} 
\def\grzweifunfzig{\affiliation{ Department of Physics, McGill University, Montr\'eal, Qu\'ebec, Canada H3A 2T8 }} 
\def\grdreifunfzig{\affiliation{Institute for Theoretical and Experimental Physics, Moscow, Russia }} 
\def\grvierfunfzig{\affiliation{Lebedev Physical Institute, Moscow, Russia }} 
\def\grfunffunfzig{\affiliation{ Moscow Engineering Physics Institute, Moscow, Russia }} 
\def\grsechsfunfzig{\affiliation{ Moscow State University, Institute of Nuclear Physics, Moscow, Russia }} 
\def\grsiebenfunfzig{\affiliation{ Max-Planck-Institut f\"ur Physik, M\"unchen, Germany }} 
\def\grachtfunfzig{\affiliation{LAL, Univ.~Paris-Sud, CNRS/IN2P3, Orsay, France }} 
\def\grneunfunfzig{\affiliation{ Department of Physics, University of Oxford, Oxford, United Kingdom }} 
\def\grsechzig{\affiliation{ INFN Padova, Padova, Italy }} 
\def\greinsechzig{\affiliation{ Dipartimento di Fisica dell'Universit\`a and INFN, Padova, Italy }} 
\def\grzweisechzig{\affiliation{LLR, Ecole Polytechnique, CNRS/IN2P3, Palaiseau, France }} 
\def\grdreisechzig{\affiliation{LPNHE, Universit\'es Paris VI and VII, CNRS/IN2P3, Paris, France }} 
\def\grviersechzig{\affiliation{Faculty of Science, University of Montenegro, Podgorica, Montenegro }} 
\def\grfunfsechzig{\affiliation{ Institute of Physics, Academy of Sciences of the Czech Republic, Praha, Czech Republic }} 
\def\grsechssechzig{\affiliation{Faculty of Mathematics and Physics, Charles University, Praha, Czech Republic }} 
\def\grsiebensechzig{\affiliation{ Department of Particle Physics, Weizmann Institute, Rehovot, Israel }} 
\def\grachtsechzig{\affiliation{Dipartimento di Fisica Universit\`a di Roma Tre and INFN Roma~3, Roma, Italy }} 
\def\grneunsechzig{\affiliation{ Dipartimento di Fisica, Universit\`a 'La Sapienza' and INFN, Rome, Italy }} 
\def\grsiebzig{\affiliation{ Polytechnic University, Sagamihara, Japan }} 
\def\greinsiebzig{\affiliation{ for Nuclear Research and Nuclear Energy, Sofia, Bulgaria }} 
\def\grzweisiebzig{\affiliation{ Raymond and Beverly Sackler Faculty of Exact Sciences, School of Physics, Tel Aviv University, Tel Aviv, Israel }} 
\def\grdreisiebzig{\affiliation{ Department of Physics, Tokyo Institute of Technology, Tokyo, Japan }} 
\def\grviersiebzig{\affiliation{ Department of Physics, University of Tokyo, Tokyo, Japan }} 
\def\grfunfsiebzig{\affiliation{ Tokyo Metropolitan University, Department of Physics, Tokyo, Japan }} 
\def\grsechssiebzig{\affiliation{ Universit\`a di Torino and INFN, Torino, Italy }} 
\def\grsiebensiebzig{\affiliation{ Universit\`a del Piemonte Orientale, Novara, and INFN, Torino, Italy }} 
\def\grachtsiebzig{\affiliation{ Department of Physics, University of Toronto, Toronto, Ontario, Canada M5S 1A7 }} 
\def\grneunsiebzig{\affiliation{ Institute of Particle and Nuclear Studies, KEK, Tsukuba, Japan }} 
\def\grachtzig{\affiliation{ Institute of Physics and Technology of the Mongolian Academy of Sciences, Ulaanbaatar, Mongolia }} 
\def\greinachtzig{\affiliation{ Department of Physics, Pennsylvania State University, University Park, Pennsylvania 16802, USA }} 
\def\grzweiachtzig{\affiliation{Paul Scherrer Institut, Villigen, Switzerland }} 
\def\grdreiachtzig{\affiliation{ Warsaw University, Institute of Experimental Physics, Warsaw, Poland }} 
\def\grvierachtzig{\affiliation{ Institute for Nuclear Studies, Warsaw, Poland }} 
\def\grfunfachtzig{\affiliation{Fachbereich C, Universit\"at Wuppertal, Wuppertal, Germany }} 
\def\grsechsachtzig{\affiliation{Yerevan Physics Institute, Yerevan, Armenia }} 
\def\grsiebenachtzig{\affiliation{ Meiji Gakuin University, Faculty of General Education, Yokohama, Japan }} 
\def\grachtachtzig{\affiliation{ Department of Physics, York University, Ontario, Canada M3J1P3 }} 
\def\grneunachtzig{\affiliation{ Deutsches Elektronen-Synchrotron DESY, Zeuthen, Germany }} 
\def\grneunzig{\affiliation{Institut f\"ur Teilchenphysik, ETH, Z\"urich, Switzerland }} 
\def\greinneunzig{\affiliation{Physik-Institut der Universit\"at Z\"urich, Z\"urich, Switzerland }} 
\def\grzweineunzig{\affiliation{ Louisiana Tech University, Ruston, Louisiana 71272, USA }} 
\def\grdreineunzig{\affiliation{ Laboratório de Física Experimental de Partículas, Av. Prof. Gama Pinto 2, 1649-003 Lisboa, Portugal }} 
\def\grvierneunzig{\affiliation{ Department of Physics, Yale University, New Haven, Connecticut 06520-8121, USA }} 
\def\grfunfneunzig{\affiliation{ School of Physics and Astronomy, University of Glasgow, Glasgow G12 8QQ, United Kingdom }} 

\def\groupalberta{\affiliation{Department of Physics, University of Alberta, Edmonton, Alberta T6G 2J1, Canada}}
\def\groupbari{\affiliation{Istituto Nazionale di Fisica Nucleare, Sezione di Bari, 70124 Bari, Italy}}
\def\groupbeijing{\affiliation{School of Physics, Peking University, Beijing 100871, China}}
\def\groupchina{\affiliation{Department of Modern Physics, University of Science and Technology of China, Hefei, Anhui 230026, China}}
\def\groupcolorado{\affiliation{Nuclear Physics Laboratory, University of Colorado, Boulder, Colorado 80309-0446, USA}}
\def\grouperlangen{\affiliation{Physikalisches Institut, Universit\"at Erlangen-N\"urnberg, 91058 Erlangen, Germany}}
\def\groupferrara{\affiliation{Istituto Nazionale di Fisica Nucleare, Sezione di Ferrara and Dipartimento di Fisica, Universit\`a di Ferrara, 44100 Ferrara, Italy}}
\def\groupfrascati{\affiliation{Istituto Nazionale di Fisica Nucleare, Laboratori Nazionali di Frascati, 00044 Frascati, Italy}}
\def\groupgent{\affiliation{Department of Subatomic and Radiation Physics, University of Gent, 9000 Gent, Belgium}}
\def\groupgiessen{\affiliation{Physikalisches Institut, Universit\"at Gie{\ss}en, 35392 Gie{\ss}en, Germany}}
\def\groupillinois{\affiliation{Department of Physics, University of Illinois, Urbana, Illinois 61801-3080, USA}}
\def\groupmit{\affiliation{Laboratory for Nuclear Science, Massachusetts Institute of Technology, Cambridge, Massachusetts 02139, USA}}
\def\groupmichigan{\affiliation{Randall Laboratory of Physics, University of Michigan, Ann Arbor, Michigan 48109-1120, USA }}
\def\groupmoscow{\affiliation{Lebedev Physical Institute, 117924 Moscow, Russia}}
\def\groupstpetersburg{\affiliation{Petersburg Nuclear Physics Institute, St. Petersburg, Gatchina, 188350 Russia}}
\def\groupprotvino{\affiliation{Institute for High Energy Physics, Protvino, Moscow region, 142281 Russia}}
\def\groupregensburg{\affiliation{Institut f\"ur Theoretische Physik, Universit\"at Regensburg, 93040 Regensburg, Germany}}
\def\grouprome{\affiliation{Istituto Nazionale di Fisica Nucleare, Sezione Roma 1, Gruppo Sanit\`a and Physics Laboratory, Istituto Superiore di Sanit\`a, 00161 Roma, Italy}}
\def\groupsimonfraser{\affiliation{Department of Physics, Simon Fraser University, Burnaby, British Columbia V5A 1S6, Canada}}
\def\grouptriumf{\affiliation{TRIUMF, Vancouver, British Columbia V6T 2A3, Canada}}
\def\groupamsterdam{\affiliation{Department of Physics and Astronomy, Vrije Universiteit, 1081 HV Amsterdam, The Netherlands}}

\author{V.~Gharibyan} \greindreisig
\author{V.~Adler} \greindreisig
\author{P.D.~Allfrey} \grneunfunfzig
\author{M.A.~Bell} \grneunfunfzig
\author{B.D.~Belusic} \greindreisig
\author{A.~Block} \greindreisig
\author{Y.~Bozhko} \greindreisig
\author{J.A.~Coughlan} \grzweizwanzig
\author{R.K.~Dementiev} \grsechsfunfzig
\author{A.~Deshpande} \grvierneunzig
\author{J.~Ferencei} \grsiebendreisig
\author{R.~Gon\c{c}alo} \grdreineunzig
\author{K.H.~Hiller} \grneunachtzig
\author{R.~Kaiser} \grneunzwanzig
\author{R.~Kammering} \greindreisig
\author{B.~Krause} \greindreisig
\author{B.-Q.~Ma} \groupbeijing
\author{M.C.K.~Mattingly} \grsechs
\author{S.~Padhi} \grzweifunfzig
\author{A.~Perieanu} \grdreisig
\author{D.~Protopopescu} \grfunfneunzig
\author{P.~Ryan} \grneunvierzig
\author{J.~Schaffran} \greindreisig
\author{P.~Schmid} \greindreisig
\author{M.~Seebach} \greindreisig
\author{E.~Sombrowski} \greindreisig
\author{G.~Susinno} \grsiebzehn
\author{M.~Wang} \grzehn
\author{K.~Wick} \grdreisig
\author{M.~Wobisch} \grzweineunzig
\author{A.~Zichichi} \grneun

%% file: gpv-rmcomment.bbl
\begin{thebibliography}{00}
\bibitem{Einstein-GR} 
  A.~Einstein,
  Annalen Phys.\  {\bf 49}, 769 (1916).

\bibitem{Will:2014kxa} 
  C.~M.~Will,
  Living Rev.\ Rel.\  {\bf 17}, 4 (2014)
  doi:10.12942/lrr-2014-4
  [arXiv:1403.7377 [gr-qc]].

\bibitem{Patrignani:2016xqp} 
  C.~Patrignani {\it et al.} [Particle Data Group],
  Chin.\ Phys.\ C {\bf 40}, no.~10, 100001 (2016).
  doi:10.1088/1674-1137/40/10/100001

\bibitem{preprintVG:2016} 
  V.~Gharibyan,
  %``Experimental Hint for Gravitational CP Violation,''
  Mod.\ Phys.\ Lett.\ A {\bf 35}, no. 11, 2050079 (2020).
  doi:10.1142/S0217732320500790 

\bibitem{Heckel:2008hw}
  B.~R.~Heckel, E.~G.~Adelberger, C.~E.~Cramer, T.~S.~Cook, S.~Schlamminger and U.~Schmidt,
 %  ``Preferred-Frame and CP-Violation Tests with Polarized Electrons,''
  Phys.\ Rev.\ D {\bf 78} (2008) 092006
  [arXiv:0808.2673 [hep-ex]].


\bibitem{Kostelecky:2004pd}
  V.~A.~Kostelecky,
 %  ``Gravity, Lorentz violation, and the standard model,''
  Phys.\ Rev.\ D {\bf 69} (2004) 105009

\bibitem{Hehl:1976kj}
  F.~W.~Hehl, P.~Von Der Heyde, G.~D.~Kerlick and J.~M.~Nester,
 %  ``General Relativity with Spin and Torsion: Foundations and Prospects,''
  Rev.\ Mod.\ Phys.\  {\bf 48} (1976) 393.

\bibitem{Moody:84}J.~E.~Moody and F.~Wilczek,
Phys.\ Rev.\ D {\bf 30} 130 (1984).

\bibitem{Ni:2009fg}
  W.~T.~Ni,
 %  ``Searches for the role of spin and polarization in gravity,''
  Rept.\ Prog.\ Phys.\  {\bf 73} (2010) 056901
  [arXiv:0912.5057 [gr-qc]].


\bibitem{Gleiser:2001rm}
  R.~J.~Gleiser and C.~N.~Kozameh,
  Phys.\ Rev.\  D {\bf 64}, 083007 (2001).

\bibitem{Altschul:2011ab} 
  B.~Altschul and M.~Mewes,
  %``Bounds on Parity Violation In the Cosmological Redshift,''
  Phys.\ Rev.\ D {\bf 84}, 083512 (2011)
  doi:10.1103/PhysRevD.84.083512
  %%CITATION = doi:10.1103/PhysRevD.84.083512;%%
  %1 citations counted in INSPIRE as of 04 Jul 2019

\bibitem{Stecker:2011ps}
  Stecker,F.W.~
  Astropart.\ Phys.\  {\bf 35} (2011) 95.

\bibitem{Evans:2001hy} 
  J.~C.~Evans, P.~M.~Alsing, S.~Giorgetti and K.~K.~Nandi,
 %  ``Matter waves in a gravitational field: An Index of refraction for massive particles in general relativity,''
  Am.\ J.\ Phys.\  {\bf 69}, 1103 (2001)
  [gr-qc/0107063].

\bibitem{deFelice:1971ui} 
  F.~de Felice,
  %``On the Gravitational field acting as an optical medium,''
  Gen.\ Rel.\ Grav.\  {\bf 2}, 347 (1971).
  %%CITATION = GRGVA,2,347;%%
  %24 citations counted in INSPIRE as of 31 Oct 2013

\bibitem{Sen:2010zzf} 
  A.~K.~Sen,
  %``A More exact expression for the gravitational deflection of light, derived using material medium approach,''
  Astrophysics {\bf 53}, 560 (2010).

\bibitem{Chu:2010tc} 
  Y.~Z.~Chu, D.~M.~Jacobs, Y.~Ng and G.~D.~Starkman,
  %``It's Hard to Learn How Gravity and Electromagnetism Couple,''
  Phys.\ Rev.\ D {\bf 82}, 064022 (2010)
  doi:10.1103/PhysRevD.82.064022

\bibitem{Pound:1960zz} 
  R.~V.~Pound and G.~A.~Rebka,
  Phys.\ Rev.\ Lett.\  {\bf 3}, 439 (1959).
  doi:10.1103/PhysRevLett.3.439

\bibitem{mcmaster:1961xe}
   W.~H.~McMaster,
 %   ``Matrix Representation of Polarization.''
   Rev.\ Mod.\ Phys.
{\bf 33} (1961) 8.

\bibitem{Kotkin:2002ra} 
  G.~L.~Kotkin, V.~G.~Serbo and V.~I.~Telnov,
 %  ``Electron (positron) beam polarization by Compton scattering on circular polarized laser photons,''
  Phys.\ Rev.\ ST Accel.\ Beams {\bf 6}, 011001 (2003)
  [hep-ph/0205139].

\bibitem{Lipps}
F.~W.~Lipps, H.~A.~Tolhoek,
Physica {\bf 20} (1954) 85, 385.

\bibitem{Anthony:2003ub} 
  P.~L.~Anthony {\it et al.} [SLAC E158 Collaboration],
  Phys.\ Rev.\ Lett.\  {\bf 92}, 181602 (2004)
  doi:10.1103/PhysRevLett.92.181602
  [hep-ex/0312035].

\bibitem{Anthony:2005pm}
P.~L.~Anthony {\it et al.}  [SLAC E158 Collaboration],
  Phys.\ Rev.\ Lett.\  {\bf 95} (2005) 081601.

\bibitem{Maas:2005fk} 
  F.~E.~Maas [A4 Collaboration],
 %  Parity violating electron scattering at the MAMI facility in Mainz,
  Eur.\ Phys.\ J.\ A {\bf 24S2}, 47 (2005).

\bibitem{Maas:2008zzc} 
  F.~E.~Maas [A4 Collaboration],
 %  Parity violating electron scattering and strangeness in the nucleon,
  AIP Conf.\ Proc.\  {\bf 1056}, 98 (2008).


\bibitem{Baudrand:2010hp} 
  S.~Baudrand {\it et al.},
 %  ``A High Precision Fabry--Perot Cavity Polarimeter at HERA,''
  JINST {\bf 5}, P06005 (2010)
  doi:10.1088/1748-0221/5/06/P06005
  [arXiv:1005.2741 [physics.ins-det]].

\bibitem{Barber:1992fc}
D.~P.~Barber {\it et al.},
Nucl.\ Instrum.\ Meth.\ A {\bf 329}, 79 (1993).

\bibitem{Lomperski:1993aw}
M.~Lomperski,
DESY-93-045

\bibitem{Gharibyan:2003fe}
V. Gharibyan,
  Phys.\ Lett.\ B {\bf 611} (2005) 231.

 % B.~Sobloher, R.~Fabbri, T.~Behnke, J.~Olsson, D.~Pitzl, S.~Schmitt and J.~Tomaszewska,
 %  ``Polarisation at HERA---Reanalysis of the HERA II Polarimeter Data,''
 % arXiv:1201.2894 [physics.ins-det].

\bibitem{James:1975dr} 
  F.~James and M.~Roos,
 %  ``Minuit: A System for Function Minimization and Analysis of the Parameter Errors and Correlations,''
  Comput.\ Phys.\ Commun.\  {\bf 10}, 343 (1975).
  doi:10.1016/0010-4655(75)90039-9

\bibitem{Kostelecky:2003fs} 
  V.~A.~Kostelecky,
  %``Gravity, Lorentz violation, and the standard model,''
  Phys.\ Rev.\ D {\bf 69}, 105009 (2004)
  doi:10.1103/PhysRevD.69.105009
  [hep-th/0312310].

\bibitem{Bailey:2006fd} 
  Q.~G.~Bailey and V.~A.~Kostelecky,
  %``Signals for Lorentz violation in post-Newtonian gravity,''
  Phys.\ Rev.\ D {\bf 74}, 045001 (2006)
  doi:10.1103/PhysRevD.74.045001
  [gr-qc/0603030].
  %%CITATION = doi:10.1103/PhysRevD.74.045001;%%
  %290 citations counted in INSPIRE as of 09 Jul 2019

\bibitem{Jackiw:2003pm} 
  R.~Jackiw and S.~Y.~Pi,
  %``Chern-Simons modification of general relativity,''
  Phys.\ Rev.\ D {\bf 68}, 104012 (2003)
  doi:10.1103/PhysRevD.68.104012
  [gr-qc/0308071].
  %%CITATION = doi:10.1103/PhysRevD.68.104012;%%
  %373 citations counted in INSPIRE as of 09 Jul 2019

\bibitem{Smith:2007jm} 
  T.~L.~Smith, A.~L.~Erickcek, R.~R.~Caldwell and M.~Kamionkowski,
  %``The Effects of Chern-Simons gravity on bodies orbiting the Earth,''
  Phys.\ Rev.\ D {\bf 77}, 024015 (2008)
  doi:10.1103/PhysRevD.77.024015
  [arXiv:0708.0001 [astro-ph]].
  %%CITATION = doi:10.1103/PhysRevD.77.024015;%%
  %100 citations counted in INSPIRE as of 09 Jul 2019

\bibitem{Bocquet:2010ke}
  J.-P.~Bocquet {\it et al.},
  %``Limits on light-speed anisotropies from Compton scattering of high-energy electrons,''
  Phys.\ Rev.\ Lett.\  {\bf 104} (2010) 241601
  [arXiv:1005.5230 [hep-ex]].
  %%CITATION = ARXIV:1005.5230;%%
  %51 citations counted in INSPIRE as of 17 Aug 2015

\bibitem{Colladay:1998fq} 
  D.~Colladay and V.~A.~Kostelecky,
  %``Lorentz violating extension of the standard model,''
  Phys.\ Rev.\ D {\bf 58}, 116002 (1998).

\bibitem{Wald:1972sz} 
  R.~M.~Wald,
  %``Gravitational spin interaction,''
  Phys.\ Rev.\ D {\bf 6}, 406 (1972).
  doi:10.1103/PhysRevD.6.406

\bibitem{Papapetrou:1951pa} 
  A.~Papapetrou,
  %``Spinning test particles in general relativity. 1.,''
  Proc.\ Roy.\ Soc.\ Lond.\ A {\bf 209}, 248 (1951).
  doi:10.1098/rspa.1951.0200

\bibitem{Bonnor:2002zg}
  W.~B.~Bonnor,
  %``Classical gravitational spin spin interaction,''
  Class.\ Quant.\ Grav.\  {\bf 19} (2002) 143
  doi:10.1088/0264-9381/19/1/308

\bibitem{Obukhov:2000ih} 
  Y.~N.~Obukhov,
  %``Spin, gravity, and inertia,''
  Phys.\ Rev.\ Lett.\  {\bf 86}, 192 (2001)
  doi:10.1103/PhysRevLett.86.192

\bibitem{Latorre:1995cv}
J.~I.~Latorre, P.~Pascual and R.~Tarrach,
Nucl.\ Phys.\ B {\bf 437}, 60 (1995)
[hep-th/9408016].

\bibitem{Dittrich:1998fy}
W.~Dittrich and H.~Gies,
Phys.\ Rev.\ D {\bf 58}, 025004 (1998)
[hep-ph/9804375].

\bibitem{Gharibyan:2006note}
  V.~Gharibyan and S.~Schmitt,
 %  ``Transverse Polarimeter Systematic Errors,''
HERMES Internal Report 06-104, (2006),\\
\url{http://www-hermes.desy.de/notes/pub/06-LIB/vahag.06-104.tpol-syserr.pdf}.

\bibitem{priv:Brinkmann}
  R.~Brinkmann, DESY, private communication, 2018.
 
\bibitem{Persic:1995ru} 
  M.~Persic, P.~Salucci and F.~Stel,
 %  ``The Universal rotation curve of spiral galaxies: 1.  The dark matter connection,''
  Mon.\ Not.\ Roy.\ Astron.\ Soc.\  {\bf 281}, 27 (1996)
  doi:10.1093/mnras/281.1.27, 10.1093/mnras/278.1.27
  [astro-ph/9506004].


\end{thebibliography}
